\documentclass[iop]{emulateapj}
\usepackage{natbib,graphicx,color}
\catcode`\@=11
\def\gsim{\ifmmode{\mathrel{\mathpalette\@versim>}}
    \else{$\mathrel{\mathpalette\@versim>}$}\fi}
\def\lsim{\ifmmode{\mathrel{\mathpalette\@versim<}}
    \else{$\mathrel{\mathpalette\@versim<}$}\fi}
\def\@versim#1#2{\lower 2.9truept \vbox{\baselineskip 0pt \lineskip
    0.5truept \ialign{$\m@th#1\hfil##\hfil$\crcr#2\crcr\sim\crcr}}}
\catcode`\@=12

\newcommand{\msun}{${\rm M_{\sun}}$}

\def\ltsima{$\; \buildrel < \over \sim \;$}
\def\simlt{\lower.5ex\hbox{\ltsima}}
\def\gtsima{$\; \buildrel > \over \sim \;$}
\def\simgt{\lower.5ex\hbox{\gtsima}}
\def\kms{{\rm\,km\,s^{-1}}}

\def\kpc{{\rm\,kpc}}

\def\msun{{\rm\,M_\odot}}


\def\deg{^\circ}

\def\s{\ifmmode \widetilde \else \~\fi}
\def\={\overline}

\def\spose#1{\hbox to 0pt{#1\hss}}

\def\etal{{\it et al.\ }}

\def\lta{\mathrel{\spose{\lower 3pt\hbox{$\mathchar"218$}}
     \raise 2.0pt\hbox{$\mathchar"13C$}}}
\def\gta{\mathrel{\spose{\lower 3pt\hbox{$\mathchar"218$}}
     \raise 2.0pt\hbox{$\mathchar"13E$}}}
\def\Dt{\spose{\raise 1.5ex\hbox{\hskip3pt$\mathchar"201$}}}    
\def\dt{\spose{\raise 1.0ex\hbox{\hskip2pt$\mathchar"201$}}}    

\def\dotsfill{\leaders\hbox to 1em{\hss.\hss}\hfill}

\def\Gyr{{\rm\,Gyr}}

\def\etal{{et al.~}}
\def\ltsima{$\; \buildrel < \over \sim \;$}
\def\gtsima{$\; \buildrel > \over \sim \;$}
\def\lsim{\lower.5ex\hbox{\ltsima}}
\def\gsim{\lower.5ex\hbox{\gtsima}}
\def\lapp{\ifmmode\stackrel{<}{_{\sim}}\else$\stackrel{<}{_{\sim}}$\fi}
\def\gapp{\ifmmode\stackrel{>}{_{\sim}}\else$\stackrel{<}{_{\sim}}$\fi}

\shorttitle{Is the Milky Way halo triaxial?}
\shortauthors{Ibata et al.}

\begin{document}

\title{Does the Sagittarius Stream constrain the Milky Way halo to be triaxial?}

\author{R. Ibata\altaffilmark{1}, G. F. Lewis\altaffilmark{2}, N. F. Martin\altaffilmark{1,3}, M. Bellazzini\altaffilmark{4}, M. Correnti\altaffilmark{5}}

\altaffiltext{1}{Observatoire Astronomique, Universit\'e de Strasbourg, CNRS, 11, rue de l'Universit\'e, F-67000 Strasbourg, France}

\altaffiltext{2}{Sydney Institute for Astronomy, School of Physics, A28, The University of Sydney, NSW 2006, Australia}

\altaffiltext{3}{Max-Planck-Institut f\"{u}r Astronomie, K\"{o}nigstuhl 17, 69117 Heidelberg, Germany}

\altaffiltext{4}{INAF-Bologna Astronomical Observatory, via Ranzani 1, I-40127 Bologna, Italy}

\altaffiltext{5}{INAF-Istituto di Astrofisica Spaziale e Fisica Cosmica, Via P. Gobetti, 101, I-40129 Bologna, Italy}

\begin{abstract}
Recent analyses of the stellar stream of the Sagittarius dwarf galaxy have claimed that the kinematics and three-dimensional
location of the M-giant stars in this structure constrain the dark matter halo of our Galaxy to possess a triaxial
shape that is extremely flattened, being essentially an oblate ellipsoid oriented perpendicular to the Galactic disk.
Using a new stream-fitting algorithm, based on a Markov Chain Monte Carlo procedure, we investigate whether this
claim remains valid if we allow the density profile of the Milky Way halo greater freedom. 
We find stream solutions that fit the leading and trailing arms of this structure even in a spherical halo, although this 
would need a rising Galactic rotation curve at large Galactocentric radius. However, the required
rotation curve is not ruled out by current constraints. It appears
therefore that for the Milky Way, halo triaxiality, despite its strong theoretical motivation, is not 
required to explain the Sagittarius stream. This degeneracy between triaxiality and the halo density profile
suggests that in future endeavors to model this structure, it will be advantageous to relax the strict analytic density 
profiles that have been used to date.
\end{abstract}


\section{Introduction}
\label{sec:int}

In standard $\Lambda$ Cold Dark Matter ($\Lambda$-CDM) cosmology \citep{2011ApJS..192...18K}
galaxies form within dark matter halos that coalesce through repeated mergers.
High resolution pure CDM simulations give rise
to significantly triaxial galaxy halos, that become progressively less triaxial towards lower sizes and masses due to the
increased dynamical age of the lower mass structures \citep{2006MNRAS.367.1781A}.
The inclusion of baryons in galaxy formation simulations alters dramatically the behavior of the central 
regions of the halo, which added to the effect of merging satellites makes the central halo become 
rounder \citep{2008ApJ...681.1076D}. 

In the Milky Way, we have perhaps the best opportunity to constrain the shape of any
dark matter distribution. One particularly promising strategy to accomplish this, and which is generally not
possible in extragalactic systems, is to use streams of stars to probe the global potential.
The streams of low-mass satellites follow fairly closely the centre of mass orbit of their progenitor, and with
careful modeling it is possible to find a solution for the stream structure within a Galactic mass distribution.
This approach has been used in the past to model the stream of the Sagittarius (Sgr) dwarf galaxy \citep{Ibata:2001be}, 
probably the most significant accretion into the Milky Way in the last $\sim 5\Gyr$. 
Recently, \citet[][hereafter LM10]{2010ApJ...714..229L}
presented a new analysis of the spatial and kinematic structure of M-giant star members of this 
stream \citep{Majewski:2003cq,Majewski:2004df}, and showed that their data could be reproduced if
the dark halo of the Milky Way has a triaxial form with a minor to major axis ratio
of $(c/a)_\Phi=0.72$ and intermediate to major axis ratio of $(b/a)_\Phi=0.99$ (in the potential). 
The structure is therefore almost an oblate ellipsoid, but with its minor axis contained within the 
Galactic plane, which presumably induces non-circular orbits at radii where the disk is not dominant. As LM10
point out, such a halo is not natural in $\Lambda$-CDM. It is also worth noting that their model
has an extremely high degree of oblateness in density. 
These conclusions have been further reinforced by the analysis of \citet[][hereafter DW12]{2013MNRAS.428..912D} 
who in fitting the stream with orbits in a Galactic mass model,
found a similar oblate dark matter halo also oriented perpendicular to the Galactic plane.
In this contribution we aim to determine whether it is possible to reproduce the structure and kinematics of the
Sgr stream, at least as well as LM10, without invoking a triaxial Galactic halo.

\section{Modelling}

One means to model the formation of the Sgr stream would be to make a mass model of the Milky Way
and an N-body model of the Sgr dwarf, and to evolve the latter inside the potential
provided by the former using an N-body integrator. Through a judicious sampling of orbital
and structural parameters of the two bodies it would be possible to assess the triaxiality issue.
This is essentially the strategy adopted by LM10, but it has the disadvantage of being
computationally extremely costly, which greatly limits the resolution of the parameter space survey.
An alternative approach, adopted by DW12, is to follow orbits in a mass distribution, however
due to self-gravity this is not a good approximation for streams of massive satellites \citep{2011MNRAS.417..198V}.

\begin{figure*}
\vbox{
\hbox{
\includegraphics[bb= 45 35 370 370, angle=0, clip, width=8.0cm]{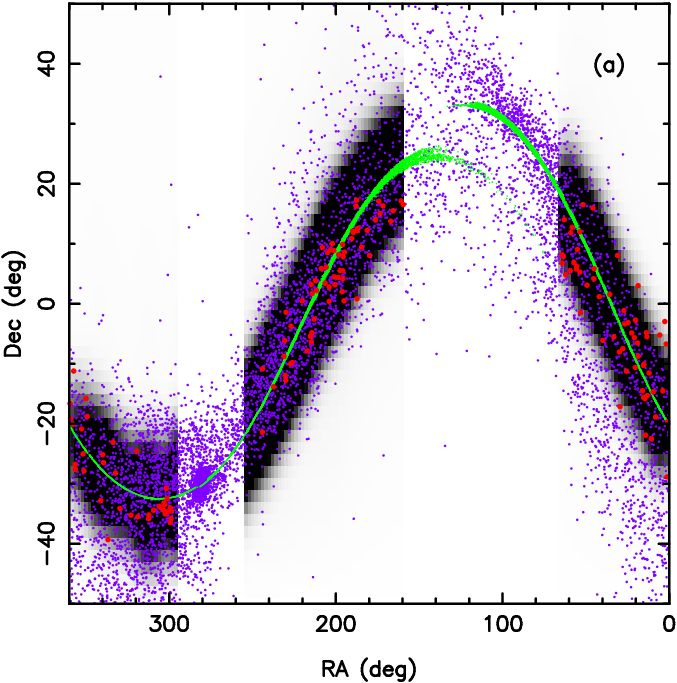}
\includegraphics[bb= 45 35 370 370, angle=0, clip, width=8.0cm]{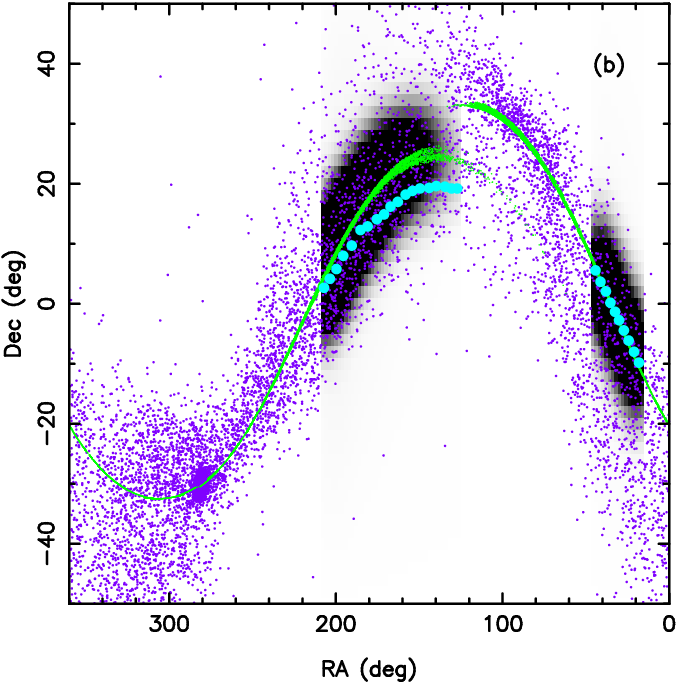}}
\hbox{
\includegraphics[bb= 45 35 370 370, angle=0, clip, width=8.0cm]{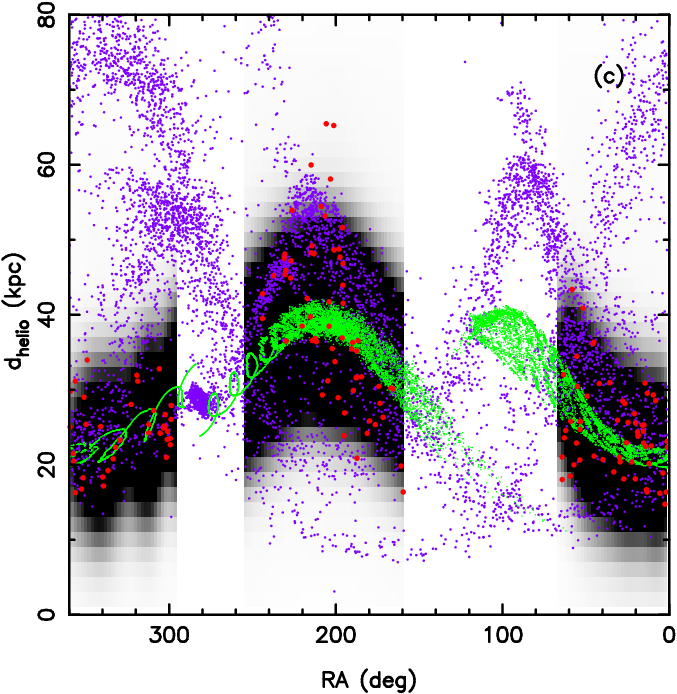}
\includegraphics[bb= 45 35 370 370, angle=0, clip, width=8.0cm]{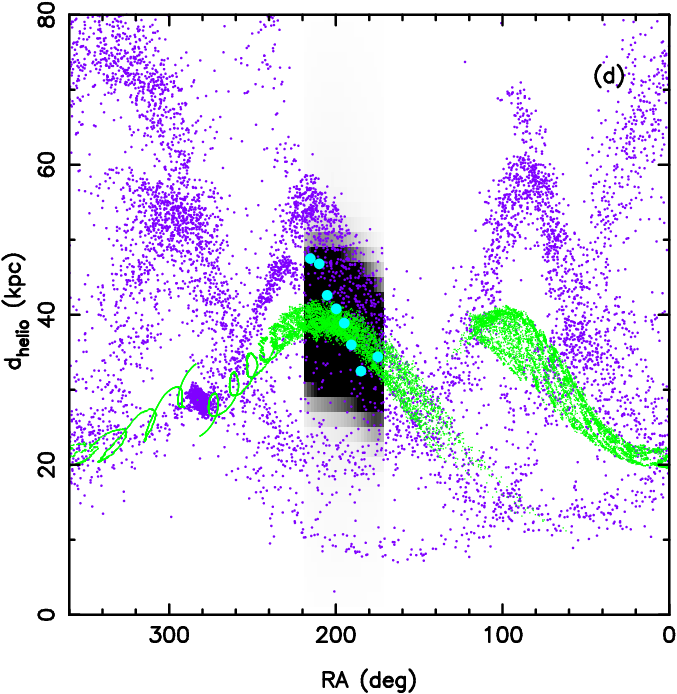}}
}
\caption{In panel (a), the sky positions in Right Ascension and Declination of the M-giant stars studied by 
LM10 (red circles) 
are plotted on top of the the best-fitting model of the Sgr stream found in $10^6$
iterations of the MCMC algorithm (green dots). The model points have been convolved with a Gaussian
function to mitigate against sampling noise in the model, resulting in the probability density
function map (black). Regions close to the Galactic plane have been masked out.
Panel (b) shows the sky regions covered by the Sloan Digital Sky Survey (SDSS), where the stream is detected:
our estimates of the SDSS sky positions of the main leading and trailing arms are shown with turquoise
points. The model points that overlap these area are shown again in green.
The lower panels show the corresponding distance profile, with (c) displaying the M-giants and (d) the 
distance estimates to red-clump stars in the SDSS as measured by Correnti et al. 2010.
The fitted model can be seen to give a good approximation to all these data. The LM10
model (a random 10\% selection of their particles) is shown in all panels with small purple dots.}
\label{fig:radec}
\end{figure*}

Instead, we have chosen to use a new stream-fitting algorithm presented in \citet{2011MNRAS.417..198V},
which searches through the parameter space of the host galaxy mass model and the 
satellite's orbital properties and mass using a Markov Chain Monte Carlo (MCMC) scheme. As we have shown
in that contribution, the method is able to provide a very close approximation to a stream
without having to resort to N-body simulations. The simple insight behind this approximation is that a leading stream
is formed from stars that leave the satellite close to the inner L1 Lagrange point, while the trailing stars
leave close to the L2 Lagrange point. In this way, we avoid having to undertake a full N-body 
simulation, and can feasibly probe millions of model parameter combinations in an automatic
way with a MCMC algorithm. 

Following \citet[][hereafter DB98]{Dehnen:1998tk}, we model the Milky Way as a sum of density 
components, using the multipole expansion technique to solve the Poisson equation. 
For the bulge, disk, thick disk and interstellar medium component, we have adopted
the parameter values of models `2b' and `2c' from DB98: these are realistic
density models that fit a wide range of observational
constraints. DB98 fit these baryonic components within a halo component modeled as
a double power law; the main difference between
their models `2b' and `2c' is that `2b' was forced to have an inner power-law exponent $\gamma_h=-2$, 
while model `2c' had  $\gamma_h=1$, similar to that of a universal (NFW) cosmological halo model 
\citep{Navarro:1997if} (naturally, these constraints on the halo affect the fitted parameters of the 
other components). However, instead of adopting the DB98 double power-law dark matter halo
model, we allow for a much larger amount of flexibility in the halo radial density profile, as we describe next.
The halo is set to be 
an axisymmetric model, defined by a density at $10\kpc$ together with an inner power law exponent $\gamma$
inside of $10\kpc$, a density at $60\kpc$ together with a power law exponent $\beta$ beyond that radius,
and density values at 8 logarithmically-spaced radial locations between 10 and $60\kpc$.
A spline function is used to interpolate the density between these anchor points.
Since the dynamics of the Sgr stream should not be very sensitive to the mass 
distribution in the inner Galaxy, or the distribution in the outer Galaxy beyond its apocenter,
we decided to fix $\gamma=1$ and $\beta=3$, similar to the inner and outer behavior of
an NFW halo model. However, the 10 halo density values
are free parameters in our modeling. Although our code also allows one to set (or fit) the flattening in density 
($q_m$) of the
axisymmetric halo component, for this particular experiment we chose to probe the case for
a spherical halo, and so kept $q_m=1$ fixed.

The remaining parameters that need to be defined are the mass, 3-dimensional distance and velocity of Sgr
as well as the position and velocity of the Sun, in order to correctly project the stream kinematics into
line of sight observables.

For the initial analysis presented here, we chose $v_{Sgr,helio}=141\kms$ \citep{Bellazzini:2008eu}, and 
following LM10 we adopted $d_{Sgr}=28\kpc$, and 
$M_{Sgr}=2.5\times10^8\msun$; these values were kept fixed. The total space velocity and direction of Sgr
was allowed to vary, as were the Galactocentric radius of the Sun $R_\odot$ and the circular velocity of the 
Local Standard of Rest $v_{LSR}$. To be consistent with \citet{Majewski:2004df}, we adopted their chosen Solar peculiar
velocity of $(u,v,w)=(-9,12,7)\kms$.

The MCMC algorithm that drives the simulation has been updated from that used in \citet{2011MNRAS.417..198V}, 
and now uses an affine-invariant ensemble-sampler scheme (with parallel tempering), identical to that 
described in \citet{Iba12}. The algorithm starts off from an initial guess for the 14 model parameters 
(10 halo density values, 2 velocity parameters for Sgr, as well as $R_\odot$ and $v_{LSR}$), to begin
its exploration of parameter space. Our initial estimations for the MCMC steps were adjusted dynamically 
in order to attain a target acceptance ratio of 25\%.

At each iteration, the algorithm makes a stream model that consists of a prediction of the locus of a tidal 
stream in space. We convert this to a probability density function by convolving the points by Gaussian functions
that approximate the combined effect of the uncertainties on the observations (position, distance, velocity) and the 
expected intrinsic width of the stream. Finally, a likelihood for this set of parameter values is calculated
from the product of the probabilities of the data (M-giant positions, distances and velocities; and SDSS field positions
and distances) and the likelihood
of the modelled Galactic rotation given the compilation by \citet{2012PASJ...64...75S}.

\section{Best-fit model}

\begin{figure}
\includegraphics[bb= 45 35 370 360, angle=0, clip, width=8.0cm]{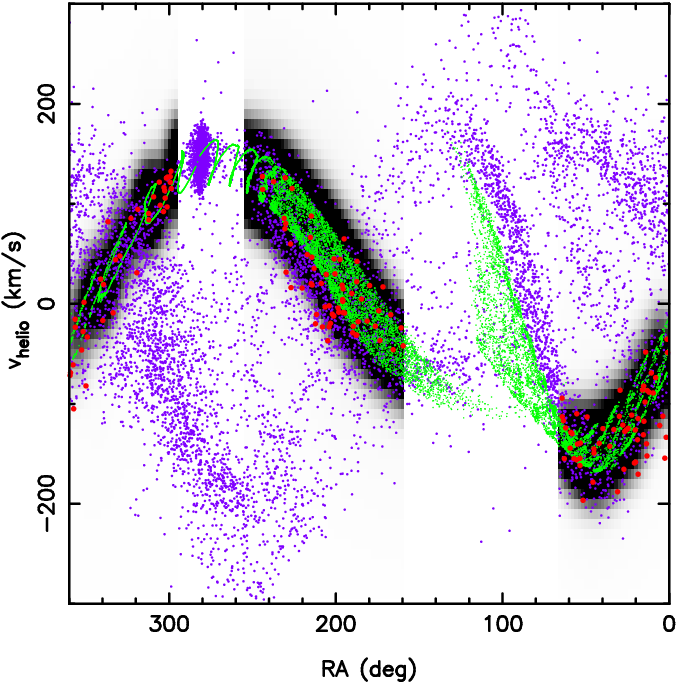}
\caption{The heliocentric velocities of the LM10 M-giants (red circles) are compared to the best-fitting model
(green points). The correspondence can be seen to be as good as that of LM10.}
\label{fig:vel}
\end{figure}

\begin{figure}
\includegraphics[bb= 45 30 660 660, angle=0, clip, width=8.0cm]{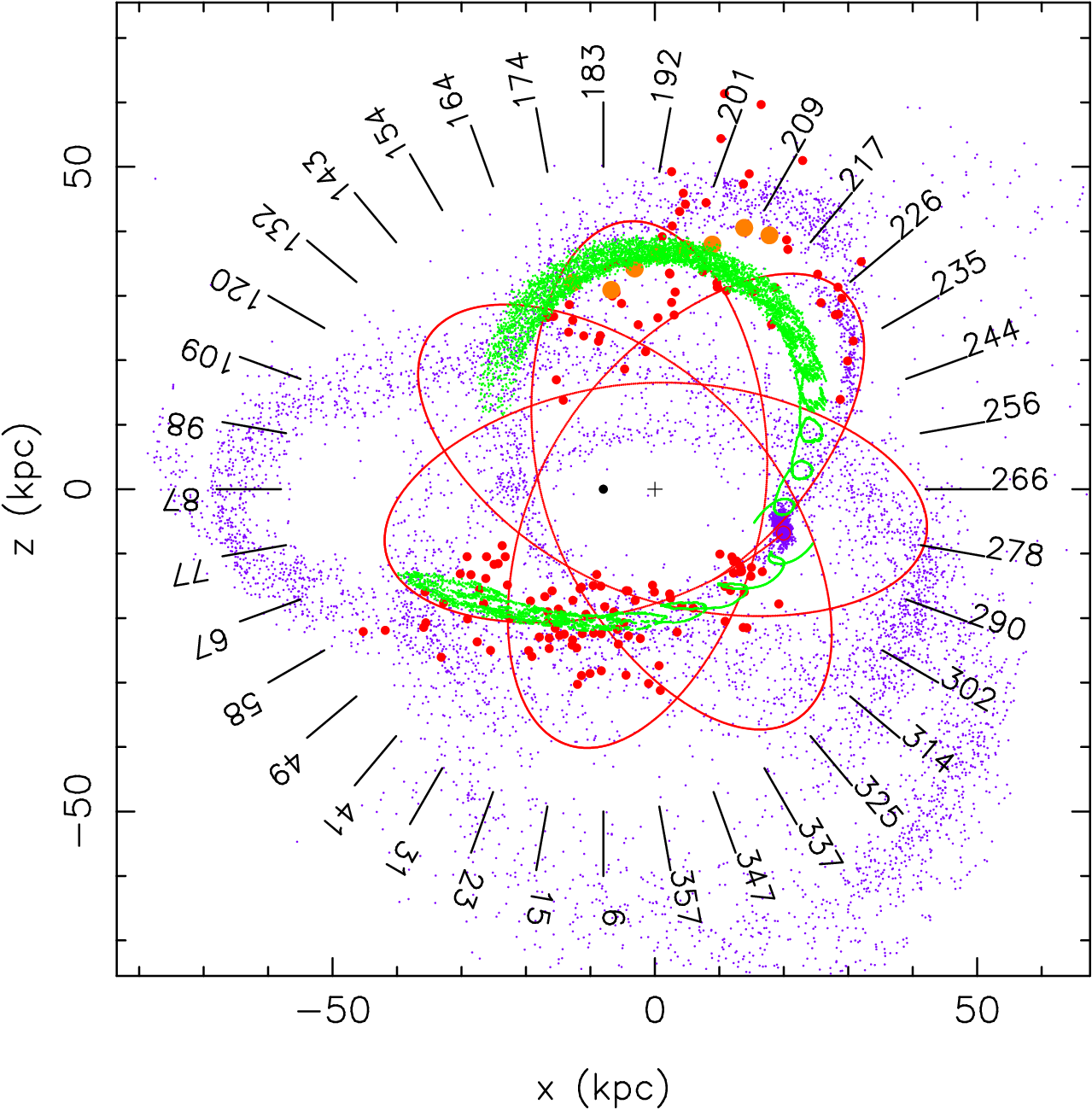}
\caption{An x-z cut through the Milky Way, showing the stream location (green dots)
according to our best-fit model. 
The path of the centre of mass of the Sgr
dwarf is shown in red; the fact that this does not line up with the stream position is due to 
the self-gravity of the dwarf galaxy, which we take into account in our modeling,
but is neglected in the DW12 analysis.
The LM10 M-giants are shown with red circles, and the big orange
dots mark the Correnti et al. 2010 `A'-field measurements from the SDSS. 
The small cross at the origin 
mark the Galactic Centre, and the position of the Sun is shown with a black circle at $(-8,0)\kpc$. The
numbers marked on the circular ticks are the corresponding Right Ascension values (in deg) in the
plane of the Sgr stream.}
\label{fig:xz}
\end{figure}

The best-fit model in $10^6$ iterations of the above algorithm adopting the 
bulge, disk, thick disk and interstellar medium parameters of model `2c' of DB98
is shown in Figures \ref{fig:radec}-\ref{fig:xz}.
Figure~\ref{fig:radec} displays the sky positions (top panels) and heliocentric distances (bottom panels) of
the MCMC solution. The locus of the stream in these parameters is displayed in green, these are convolved
with the parameter uncertainties and intrinsic stream width estimate to produce the probability density function
displayed in the background grayscale image. For comparison we also
display the LM10 triaxial N-body model (small purple dots). The left-hand panels show the
positions of M-giant stars (red circles) from LM10
within $15\deg$ of the orbital plane of the Sgr dwarf \citep{Majewski:2003cq}; RA ranges close to the disk
or close to the main body of the Sgr dwarf have been masked out. We discarded the region immediately surrounding Sgr
as we wanted to avoid being affected by the dynamical properties of the progenitor (e.g. rotation).
Panel (b) shows the sky position of the main detection of the Sgr stream in the 
SDSS \citep{Belokurov:2006ev,2012ApJ...750...80K}, as measured by us directly from SDSS data.
Panel (d) shows the distance measurements of \citet{2010ApJ...721..329C}. In both right-hand panels
we have masked out ranges in RA not probed by the data. 
It can be seen that the MCMC solution follows closely the position and distance constraints.

Figure~\ref{fig:vel} compares the radial velocity profile of the model to the M-giant data: the kinematic
behavior is similar to that of the LM10 model, and reproduces well the trend in the M-giant radial velocities.
The three-dimensional structure of the data and models is shown in Figure~\ref{fig:xz}, which gives
an edge-on view of the Milky Way. As before, the red circles are the selection of LM10 M-giants, with 
SDSS red-clump distances from \citet{2010ApJ...721..329C} shown with orange circles. The green dots mark the
locus of the MCMC stream solution: the correspondence with the M-giant distances is not perfect in the 
leading arm, but neither is that of the LM10 N-body simulation (purple dots). Note that the M-giant distance 
uncertainties are considerable ($\sim 20\%$ --- \citealt{Majewski:2003cq}), and there may also be a
systematic difference with respect to the measurements based on the red clump or main sequence.

A comparison of the relative likelihood of our model versus that of LM10 is not very useful at this stage,
the models are very different, and indeed both models fail to reproduce the data in detail
(such as the stream bifurcation \citealt{Belokurov:2006ev}), 
but it can be seen that in position, distance and velocity the
two models are similar, and give a similarly good representation of the youngest parts of the leading and trailing
arms.

While we use very similar constraints and assumptions to LM10, we have fit the Sgr stream 
without having to invoke the need for triaxiality. However, the price to pay for this is a peculiar
radial mass distribution in the halo. Figure~\ref{fig:rotationcurve} shows the required rotation curve
and corresponding density profile of the fitted spherical halo, again using the (fixed) non-halo 
components of model `2c' from DB98. The red dots in panel (a) mark the
measurements from \citet{2012PASJ...64...75S}. Beyond the Solar Circle there is substantial uncertainty
in the rotation curve, as the large scatter reveals, and beyond $20\kpc$ there are virtually no 
direct constraints. Expectations from $\Lambda$-CDM would suggest that the rotation curve
should decline slowly beyond $10\kpc$ \citep{Klypin:2002bm}, in stark contrast to our model's 
requirement. Nevertheless, we note that HI observations of our neighboring galaxy, 
Andromeda (black dots in panel `a'  --- \citealt{Chemin:2009dw}), show a rotation curve rise beyond
$25\kpc$ that is similar to our model. 

While it was not our intention to attempt to fit an accurate mass model of the inner Galaxy, it
is worth noting that the surface mass density of the disk and the vertical force $K_z$ lie within observationally-acceptable
ranges (see DB98). In addition, the Oort constants for our model (which are not identical to 
those of DB98 since we have altered the halo) are $A=13.8$ and $B=-13.5$, close to the observed
$A=14\pm 0.8$, $B=-12.4\pm 0.6$ (see Table~3 of DW12 for comparison to other solutions). 
The detailed structure of the mass distribution in the inner Galaxy turns out to be relatively unimportant in fitting
the Sgr stream: if we use the non-halo components of model `2b' from DB98 (instead of those
of `2c' as in Figures 1-4) the MCMC algorithm also converges on similar solutions that require a rising
rotation curve.

\section{Discussion and conclusions}

The analysis we have presented here shows that it is possible to reproduce the spatial and
kinematic structure of the Sgr stream at approximately the same level of precision as
LM10 without the need for a triaxial halo, if the Galactic rotation curve rises significantly in the
range $\sim 20\kpc$ to $\sim 60\kpc$. Interestingly, the break in the density profile occurs at 
approximately where the outer stellar halo component proposed by \citet{Carollo:2007fw} becomes
dominant. This rotation curve rise of course places a substantial amount of
matter in the outer Galaxy, which has the effect of braking the leading arm, bringing its velocity 
into agreement with that of the M-giants, particularly in the region around $RA \sim 200\deg$.
In addition to the MCMC parameter survey discussed above, we have also attempted an
interactive investigation, choosing parameter combinations judiciously; however, we were
unable to find another acceptable solution without a strong rotation curve rise.

\begin{figure*}
\vbox{
\centerline{
\hbox{
\includegraphics[bb= 45 30 370 360, angle=0, clip, width=8.0cm]{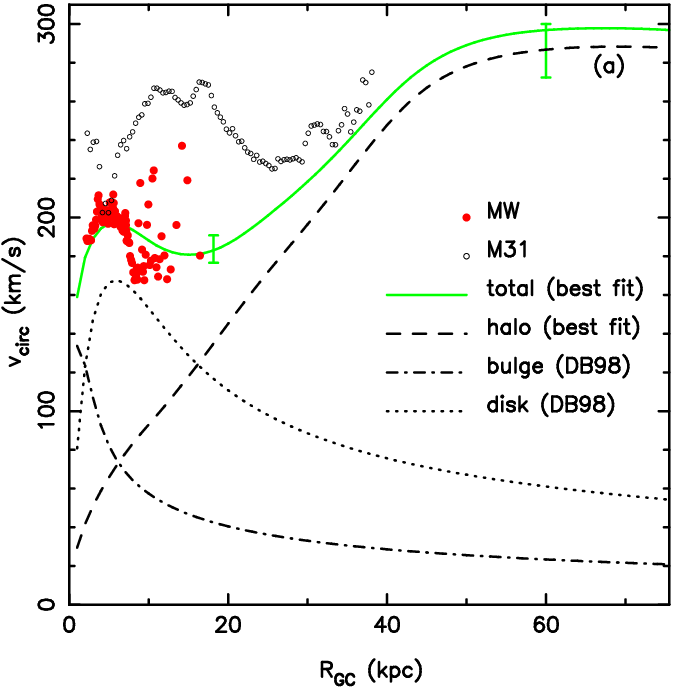}
\includegraphics[bb= 45 30 370 360, angle=0, clip, width=8.0cm]{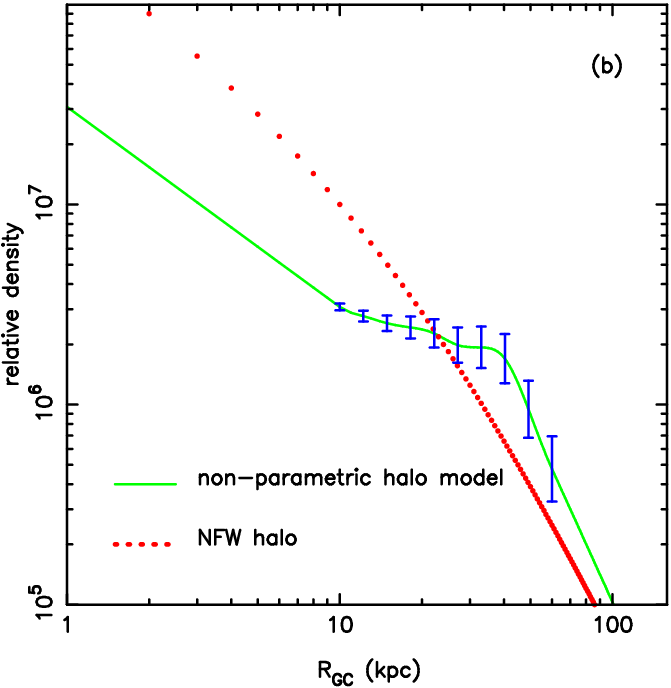}}}
}
\caption{The rotation curve required by our non-parametric spherical halo model is shown 
with a solid line in panel (a), with its decomposition into halo (dashed line), bulge
(dot-dashed line) and disk (dotted line) components.
This rotation curve fits the available measurements of \citet{2012PASJ...64...75S}
(red dots), but it requires significantly
rising values beyond $\sim 30\kpc$. Although such a rising rotation curve is unexpected, it is similar
to the observed HI rotation curve of the Andromeda galaxy (black circles --- \citealt{Chemin:2009dw}).
The two error bars at $R_{GC}=18\kpc$ and $60\kpc$ show the $1\sigma$ uncertainties derived from the MCMC
algorithm (note that a prior impedes solutions with $v_{circ} > 300\kms$).
Panel (b) shows the corresponding density profile (solid line). The error bars mark the 
radial location of the density anchor points together with the $1\sigma$ spread from the MCMC fit.
For reference we also show an NFW profile (dots) of scale radius $21.5\kpc$ \citep{Klypin:2002bm}: while this has an
identical asymptotic behavior to the halo model shown, it is substantially different in the radial range probed by the Sgr
stream. }
\label{fig:rotationcurve}
\end{figure*}

It is unclear at present 
what the structure of galactic mass dark matter halos actually is in nature, but in $\Lambda$-CDM,
dark matter density profiles decline as $r^{-2}$ only when averaged over a very large 
number of halos. The profile of Fig.~\ref{fig:rotationcurve}b is therefore not inconsistent with $\Lambda$-CDM,
and indeed a rotation curve rise of similar magnitude is observed in M31 over a similar radial range 
(Fig.~\ref{fig:rotationcurve}a).
We suspect (although we have not modeled this) that there is a trade-off between
triaxiality and rotation curve rise, so that a less triaxial model than that of LM10 
(which is actually an oblate spheroid oriented perpendicular to the Galactic disk) can be
fitted with a less extreme rotation curve than that presented here. 

A consequence of this new solution is that the Milky Way would be considerably more massive than
estimated from models with a falling rotation curve within $R<60\kpc$. Assuming a virial radius
of $258\kpc$ \citep{Klypin:2002bm}, leads to a virial mass of $2.6$--$3.1\times10^{12}\msun$ for our model.
It is interesting to note that the LMC and SMC would clearly be bound to the Milky Way if this reflects
the true mass of our Galaxy, contrary to conclusions based on HST proper motion measurements
that assume a standard Galactic mass model (see.e.g., \citealt{2007ApJ...668..949B}).
To be consistent with the timing argument limit for the mass of the Local Group of 
$5.58^{+0.26}_{-0.25}\times10^{12}\msun$ \citep{2008ApJ...678..187V}, the dark matter
would have to be primarily concentrated within the two giant galaxies. While the above mass
for the Milky Way is certainly substantially higher than most recent estimates
(e.g., \citealt{2010MNRAS.406..264W}), almost all the mass constraints at
large distance are based on Jeans equations solutions that make the unjustified assumptions
of statistical independence of the halo tracers and some form for the orbital anisotropy.

In summary, we present this solution as a means to avoid the 
triaxial halo models of LM10 or DW12, and in particular their peculiar orientation as oblate
structures perpendicular to the Galactic plane. We recognize of course that the solution presented here
is also strange, being spherical and possessing a very high mass, but we judge
that it is not ruled out by observational constraints. The significant shortcomings of both 
the LM10/DW12 and our models suggest that they
do not include sufficient of the actual complexity of (probably both)
Sgr and the Milky Way halo. Future modeling of the more ancient
parts of the Sgr stream may help discriminate between these models, 
and pave the way to better solutions.

\acknowledgments
R.A.I. gratefully acknowledges support from the Agence Nationale de la Recherche though the grant POMMME (ANR 09-BLAN-0228). GFL thanks the Australian Research Council for
support through his Future Fellowship (FT100100268) and
Discovery Project (DP110100678)

\end{document}